\documentstyle[pre,eqsecnum,aps]{revtex}
\input epsf
\begin{document}

\title{Critical point and coexistence curve properties of the
Lennard-Jones fluid: A finite-size scaling study.}

\author{Nigel B. Wilding}
\address{Institut f\"{u}r Physik, Johannes Gutenberg Universit\"{a}t, \\
Staudinger Weg 7, D-55099 Mainz, Germany.}

\date{December 1994}
\setcounter{page}{0}
\maketitle

\begin{abstract}

Monte Carlo simulations within the grand canonical ensemble are used
to explore the liquid-vapour coexistence curve and critical point
properties of the Lennard-Jones fluid. Attention is focused on the
joint distribution of density and energy fluctuations at
coexistence. In the vicinity of the critical point, this distribution
is analysed using mixed-field finite-size scaling techniques aided by
histogram reweighting methods. The analysis yields highly accurate
estimates of the critical point parameters, as well as exposing the
size and character of corrections to scaling.  In the sub-critical
coexistence region the density distribution is obtained by combining
multicanonical simulations with histogram reweighting techniques. It
is demonstrated that this procedure permits an efficient and accurate
mapping of the coexistence curve, even deep within the two phase
region.

\end{abstract}
\thispagestyle{empty}
%\twocolumn
\newpage

\section{Introduction}
\label{sec:intro}

The Lennard-Jones (LJ) fluid constitutes the prototype model for
realistic atomic fluids and has been the focus of numerous simulation
studies spanning well over 25 years
\cite{HANSEN,LEVESQUE,VERLET,AD2,AD1,NICOLAS,PANAGIO,FINN,SMIT1,SMIT,LOTFI}.
The motivation for the longstanding interest in the model is its utility
as a test-bed for ever more accurate and sophisticated theories of the
liquid state.  Contemporary theories \cite{CACCAMO,REAT1,REAT2} now
provide good agreement with simulation results over a wide range of
noncritical temperatures.  The continuing challenge, however, is to
realise a similar degree of accuracy in the critical region, where the
unbounded growth of correlations poses potentially serious
difficulties for theory and simulation alike.

One popular simulation method for studying the coexistence regime of
fluid systems is the Gibbs ensemble Monte Carlo simulation technique
(GEMC) introduced by Panagiotopoulos \cite{PANAGIO}.  In the GEMC
method, the two coexisting phases separate into two physically
detached, but thermodynamically connected boxes, the volumes of which
are allowed to fluctuate under a constant pressure environment.
Measurements of the particle density in each box provide estimates of
the coexistence densities. Common practice is to fit the temperature
dependence of these densities using a power law, the extrapolation of
which yields estimates of the critical point parameters.  The strength
of the GEMC method lies in its elimination of the physical interface
between the coexistence phases, the large free energy of which,
plagues conventional grand canonical simulations of phase coexistence
in the form of long lived metastable states and extended tunneling
times.  A large number of recent GEMC studies \cite{PANAGIO2} testify
to the method's efficacy in determining the subcritical coexistence
properties of fluids.

In the neighbourhood of the critical point, however, and for reasons
discussed in references \cite{WILDING1,MON,ALLEN,RECHT}, the GEMC
method cannot be relied upon to provide accurate estimates of the
coexistence curve parameters. Instead it is necessary to employ
finite-size scaling (FSS) techniques to probe the critical limit. FSS
techniques were originally developed in the context of computer
simulation studies of critical phenomena in spin models, and provide a
highly effective route to infinite volume critical parameters from
simulations of finite size \cite{PRIVMAN,BINDER1}. Recently their use
has been extended to fluids by explicitly incorporating the
consequences of the lack of symmetry between the coexisting phases
\cite{WILDING1,BRUCE2,WILDING2,WILDING3}.  This reduced symmetry of
fluids with respect to magnetic systems such as the Ising model, is
manifest in the so-called `field-mixing' phenomenon which is a crucial
issue in the critical behaviour of fluids. The mixed-field FSS theory
has been successfully employed in conjunction with simulations in the
grand canonical ensemble (GCE) to study the critical behaviour of a
number of critical fluid systems
\cite{WILDING1,BRUCE2,WILDING2,WILDING3,MUELLER}, including the two
dimensional LJ fluid and a three dimensional lattice model for polymer
mixtures. In the present work we extend these studies to the 3D LJ
fluid making additional use of two important new methodological
advances in computer simulation.

The essential ideas underpinning the FSS methods we shall use, are the
dual concepts of scale invariance and universality. Precisely at
criticality, the fluctuation spectra (distribution functions) of
certain readily accessible observables assume scale invariant forms
\cite{BRUCE1,BINDER2,HILFER1}.  Moreover these critical scaling
functions are {\em universal}, being identical for all members of the
same universality class. Experimental \cite{PESTAK,NARGER,SENGERS},
theoretical \cite{ZHANG,NICOLL} and simulation
\cite{BRUCE2} results show that the critical behaviour of
simple fluids corresponds to the Ising universality class (to which
all systems with short range interactions and a scalar order parameter
belong). Scaling functions measured for the critical Ising model
therefore constitute a {\em hallmark} of the Ising universality class,
a fact that can be exploited to obtain accurate estimates of the
critical point parameters of simple fluids.

Besides the progress in extending the application of FSS concepts to
fluid systems, two recent technical advances in computer simulation
methods also greatly improve the efficiency with which one can tackle
both the critical and subcritical regimes of model systems. The first
is the histogram reweighting technique of Ferrenberg and Swendsen
\cite{FERRENBERG}. This technique hinges on the observation that
histograms of observables accumulated at one set of model parameters
can be reweighted to yield estimates for histograms appropriate to
another set of parameters. The histogram reweighting method has been
found to be especially profitable close to the critical point where,
owing to the large critical fluctuations, a single simulation affords
reliable extrapolations over the entire critical region.

The second technical advance came with the introduction by Berg and
Neuhaus \cite{BERG} of the multicanonical ensemble, use of which
permits efficient Monte Carlo studies of two-phase coexistence even
far below the critical point. The multicanonical method employs a
preweighting scheme to surmount the free energy barrier associated
with formation of interfaces between the coexisting phases.  This
barrier grows rapidly as one moves further one from the critical
point, quickly rendering conventional GCE simulations impractical. The
multicanonical technique circumvents this problem by sampling not from
a Boltzmann distribution, but from a preweighted distribution that is
approximately {\em flat} between the coexisting densities. The desired
Boltzmann distributed quantities are subsequently obtained by dividing
out the preweighting factors from the measured histograms.  Use of
this method has been shown to reduce the tunnelling time to a simple
power law in the system size \cite{JANKE1}.

In the light of these new developments and the demand for ever
increasing accuracy in estimates of the phase coexistence properties
of prototype models, it seems appropriate to revisit the Lennard-Jones
fluid with a view to performing a high precision study of its critical
point and coexistence curve parameters. To this end we have carried
out a detailed simulation study of the model, bringing to bear all the
aforementioned methodological advances. Our paper is organised as
follows. We begin in section~\ref{sec:back} by providing a short
resum\'{e} of the mixed-field FSS theory for the density and energy
fluctuations of near-critical fluids. In section~\ref{sec:results}, we
present measurements of the near critical scaling operator
distributions as a function of system size. These distributions are
analysed within the FSS framework, to yield extremely accurate
estimates for the critical point and field mixing parameters.  Turning
then to the subcritical region we present the results of
multicanonical simulations measurements of the coexistence density
distributions. It is demonstrated how the measured coexistence
densities can be used in conjunction with knowledge of the critical
point parameters to construct the infinite volume coexistence
curve. Finally, in section~\ref{sec:conc} we detail our conclusions.

\section{Theoretical background}
\label{sec:back}

In this section we provide a brief overview of the principal features
of the mixed-field FSS theory of reference \cite{WILDING1}.

The system we consider is assumed to be contained in a volume $L^d$
(with $d=3$ in the simulations to be considered below) and
thermodynamically open so that the particle number can fluctuate. The
observables on which we shall focus are the particle number density:

\begin{equation}
\rho =L^{-d}N
\label{eq:phi}
\end{equation}
and the dimensionless energy density:
\begin{equation}
u=L^{-d}(4w)^{-1}\Phi(\{{\bf r}\})
\label{eq:u}
\end{equation}
where $\Phi(\{{\bf r}\})$ is the configurational energy of the system
which we assume takes the form:

\begin{equation}
\Phi(\{{\bf r}\})=\sum_{i,j}\phi(|{\bf r_i}-{\bf r_j}|) ,
\end{equation}
and where we assign the potential $\phi(r)$ the familiar Lennard-Jones form

\begin{equation}
\phi(r)=4w[(\sigma/r)^{12}-(\sigma/r)^6]\label{eq:LJdef}
\end{equation}
with $w$ the well-depth, and $\sigma$ a parameter that serves to set
the length scale.

Within the grand canonical ensemble (GCE), the joint distribution of
density and energy fluctuations, $p_L(\rho,u)$, is controlled by the
reduced chemical potential $\mu$ and the well depth $w$ (both in units
of $k_BT$).  The critical point is located by critical values of the
chemical potential $\mu_c$ and well depth $w_c$. Deviations of $w$ and
$\mu$ from their critical values control the sizes of the two relevant
scaling field that characterise the critical behaviour
\cite{WEGNER}. In the absence of the special Ising (`particle-hole')
symmetry, the relevant scaling fields comprise (asymptotically) {\em
linear combinations} of the coupling and chemical potential differences
\cite{REHR}:

\begin{equation}
\tau = w_c-w+s(\mu - \mu_c) \hspace{1cm} h=\mu - \mu_c+ r(w_c-w)
\label{eq:scaflds}
\end{equation}
where $\tau$ is the thermal scaling field and $h$ is the
ordering scaling field.  The parameters $s$ and $r$ are
system-specific quantities controlling the degree of field mixing. In
particular $r$ is identifiable as the limiting critical gradient
of the coexistence curve in
the space of $\mu$ and $w$. The role of $s$ is somewhat
less tangible; it controls the degree to which the chemical potential
features in the thermal scaling field, manifest in
the widely observed critical singularity of the coexistence curve
diameter of fluids \cite{SENGERS}.

Conjugate to the two relevant scaling fields are scaling operators
${\cal M}$ and ${\cal E}$, which comprise linear combinations of the particle
density and energy density \cite{WILDING1}:

\begin{equation}
{\cal M}  = \frac{1}{1-sr} \left[ \rho - s u \right] \hspace{1cm}
{\cal E}  =  \frac{1}{1-sr} \left[  u  - r \rho \right]
\label{eq:oplinks}
\end{equation}
The operator ${\cal M}$ (which is conjugate to the ordering field $h$) is
termed the ordering operator, while ${\cal E}$ (conjugate to the
thermal field) is termed the energy-like operator.  In the
special case of models of the Ising symmetry, (for which $s = r
=0$), ${\cal M}$ is simply the magnetisation while ${\cal E}$ is the energy
density.

The joint distribution of density and energy is simply related to the
joint distribution of mixed operators:

\begin{equation}
p_L(\rho,u) = \frac{1}{1-sr} p_L({\cal M}, {\cal E})
\label{eq:pdflink}
\end{equation}
Near criticality, and in the limit of large system size, $p_L({\cal M},{\cal
E})$
is expected to be describable by a finite-size scaling relation of the
form \cite{WILDING1}:

\begin{mathletters}
\begin{equation}
\label{eq:ansatz}
p_L({\cal M}, {\cal E}) \simeq  \Lambda _{\cal M}^+ \Lambda _{{\cal
E}}^+ \tilde{p}_{{\cal M},{\cal E}} (\Lambda _{\cal M}^+
\delta {\cal M} , \Lambda _{{\cal E}}^+ \delta {\cal E} ,
\Lambda_{\cal M} h ,
 \Lambda_{\cal E} \tau)
\end{equation}
where
\begin{equation}
\label{eq:Lamdefs}
\Lambda_{\cal E} = a_{\cal E} L^{1/\nu}, \Lambda_{\cal M} = a_{\cal
M} L^{d-\beta/\nu},\\ \Lambda_{\cal M} \Lambda _{\cal M}^+ =
\Lambda_{\cal E} \Lambda _{{\cal E}}^+ = L^d
\end{equation}
and
\begin{equation}
\label{eq:deltops}
\delta {\cal M} \equiv  {\cal M} - <{\cal M} >_c  \hspace{9mm} \delta
{\cal E} \equiv {\cal E} - <{\cal E} >_c
\end{equation}
\end{mathletters}
The subscripts c in equations~\ref{eq:deltops} signify that the averages are to
be taken at criticality.  Given appropriate choices for the non-universal
scale factors $a_{\cal M}$ and $a_{\cal E}$ (equation~\ref{eq:Lamdefs}), the
function $\tilde{p}_{{\cal M},{\cal E}}(\Lambda _{\cal M}^+ \delta
{\cal M} , \Lambda _{{\cal E}}^+ \delta {\cal E} ,
\Lambda_{\cal M} h , \Lambda_{\cal E} \tau)$ is expected to be universal.
Precisely at criticality, equation~\ref{eq:ansatz} implies simply

\begin{equation}
p_L({\cal M} , {\cal E}) \simeq  \Lambda _{\cal M}^+ \Lambda _{{\cal
E}}^+ \tilde{p}^\star_{{\cal M},{\cal E}} (\Lambda _{\cal M}^+ \delta
{\cal M} , \Lambda _{{\cal E}}^+ \delta {\cal E})
\label{eq:critlim}
\end{equation}
where $\tilde{p}_{{\cal M},{\cal E}}(x,y)=\tilde{p}_{{\cal M},
{\cal E}}(x,y,0,0)$ is a function describing the
universal and statistically scale invariant operator fluctuations
characteristic of the critical point.

For smaller system sizes, one anticipates that corrections to scaling
associated with finite values of the {\em irrelevant} scaling fields
will become significant \cite{WEGNER}. These irrelevant fields take
the form $a_1\tau^\theta+a_2\tau^{2\theta}+\ldots$, where $\theta$ is
the universal correction to scaling exponent, whose value has been
estimated to be $\theta\approx 0.54$ for the 3D Ising
class \cite{FISHER}.  Incorporating the least irrelevant of these corrections
into
equation~\ref{eq:critlim}, one finally obtains

\begin{equation}
p_L({\cal M} , {\cal E}) \simeq  \Lambda _{\cal M}^+ \Lambda _{{\cal
E}}^+ \tilde{p}^\star_{{\cal M},{\cal E}} (\Lambda _{\cal M}^+ \delta
{\cal M} ,
\Lambda _{{\cal E}}^+ \delta {\cal E}, a_1L^{-\theta/\nu})
\label{eq:critlimnew}
\end{equation}
As we shall show, it is necessary to take account of such correction
terms if highly accurate estimates of the critical parameters are to
be obtained.

\section{Monte Carlo studies}
\label{sec:results}
\subsection{Computational details}

The Monte-Carlo simulations described here were performed using a
Metropolis algorithm within the grand canonical ensemble. The
algorithm employed is similar in form to that described by Adams
\cite{AD3,ALLEN2}, but differs in the respect that only particle
transfer (insertion and deletion) steps were implemented, leaving
particle moves to be performed implicitly as a result of repeated
transfers. Physically this choice is motivated by the need to direct
the computational effort at the density fluctuations, which
are the bottleneck for phase space evolution at coexistence.

As is common practice in simulations of systems whose interparticle
potential decays rapidly with particle separation, the Lennard-Jones
potential was truncated in order to reduce the computational
effort. In accordance with most previous studies of the LJ system, the
cutoff radius was chosen to be $r_c=2.5\sigma$, and the potential was
left unshifted.  It should be noted, however, that the choice of
cutoff can have quite marked effects on the critical point parameters,
a point emphasised by Smit \cite{SMIT}.

In order to facilitate efficient computation of interparticle
interactions, the periodic simulation space of volume $L^3$ was
partitioned into $m^3$ cubic cells, each of side $r_c$. This strategy
ensures that interactions emanating from particles in a given cell
extend at most to particles in the $26$ neighbouring cells. We chose
to study a range of system sizes corresponding to $m=3,4,5,6$ and $7$,
containing at coexistence, average particle numbers of
$135,320,625,1080$ and $1715$ respectively. For the $m=3,4$ and $5$
system sizes, equilibration periods of $10^5$ Monte Carlo transfer
attempts {\em per cell} (MCS) were utilised, while for the $m=6$ and
$m=7$ system sizes up to $2\times10^6$ MCS were employed. Sampling
frequencies ranged from $15$ MCS for the $m=3$ system to $50$ MCS for
the $m=7$ system. The total length of the production runs was also
dependent upon the system size. For the $m=3$ system size,
$1\times10^7$ MCS were employed, while for the $m=7$ system, runs of
up to $5\times 10^7$ MCS were necessary. In the subcritical
coexistence region, (studied using multicanonical simulations), runs
of length $5\times10^6$ MCS were utilised. In both the sub-critical
and critical coexistence regimes, the average acceptance rate for
particle transfers was approximately $25\%$.

In the course of the simulations, the observables recorded were the
particle number density $\rho$ and the energy density $u$. The joint
distribution $p_L(\rho ,u)$ was accumulated in the form of a
histogram. In accordance with convention, we express $\rho$ and $u$ in
reduced units:

\begin{equation}
\rho^\star = \rho \sigma ^d \hspace{1cm} u^\star = u \sigma ^d
\label{eq:rhostar}
\end{equation}
We also note for future reference that the algorithm actually utilises
not the true chemical potential $\mu$ featuring in equation ~\ref{eq:scaflds},
but an effective chemical potential $\mu^\star$ to which the true
chemical potential is related by

\begin{equation}
\mu = \mu ^{\star} + \mu _0 - ln \left ( N/L^d \right ) \label{eq:mustar}
\end{equation}
where $\mu_0$ is the chemical potential in the non-interacting (ideal
gas) limit. It is this effective value that features in the results
that follow.

\subsection{The critical limit}

The most recent Gibbs ensemble simulation studies of the LJ fluid,
(using $r_c=2.5\sigma$), place the critical temperature at
$T^\star\equiv 4/w=1.176(8)$ \cite{PANAGIO3}. Using this estimate, we
attempted to locate the liquid-vapour coexistence curve by performing
a series of very short runs for the $m=4$ system size, in which the
effective chemical potential $\mu^\star$ was tuned until the density
distribution exhibited a double peaked structure. Having obtained, in
this manner, an approximate estimate of the coexistence chemical
potential, a longer run comprising $2\times10^7$ MCS was performed to
accumulate better statistics. Histogram reweighting was then applied
to the resulting histogram enabling exploration of the coexistence
curve in the neighbourhood of the simulation temperature.

To facilitate a precise identification of the coexistence chemical
potential, we adopted the criterion that the ordering operator
distribution $p_L({\cal M}) =\int d{\cal E}p_L({\cal M},{\cal E})$
must be symmetric in ${\cal M} - \langle{\cal M}\rangle$. This
criterion is the counterpart of the coexistence symmetry condition for
the Ising model magnetisation distribution. By simultaneously tuning
$\mu$ and $s$ in the reweighting of the joint distribution
$p_L(\rho,u)$, estimates for the coexistence chemical potential and
the value of the field mixing parameter $s$ that satisfy this symmetry
condition, were readily obtained.

\begin{figure}[h]
\vspace*{0.5 in}
\setlength{\epsfxsize}{19cm}
\centerline{\mbox{\epsffile{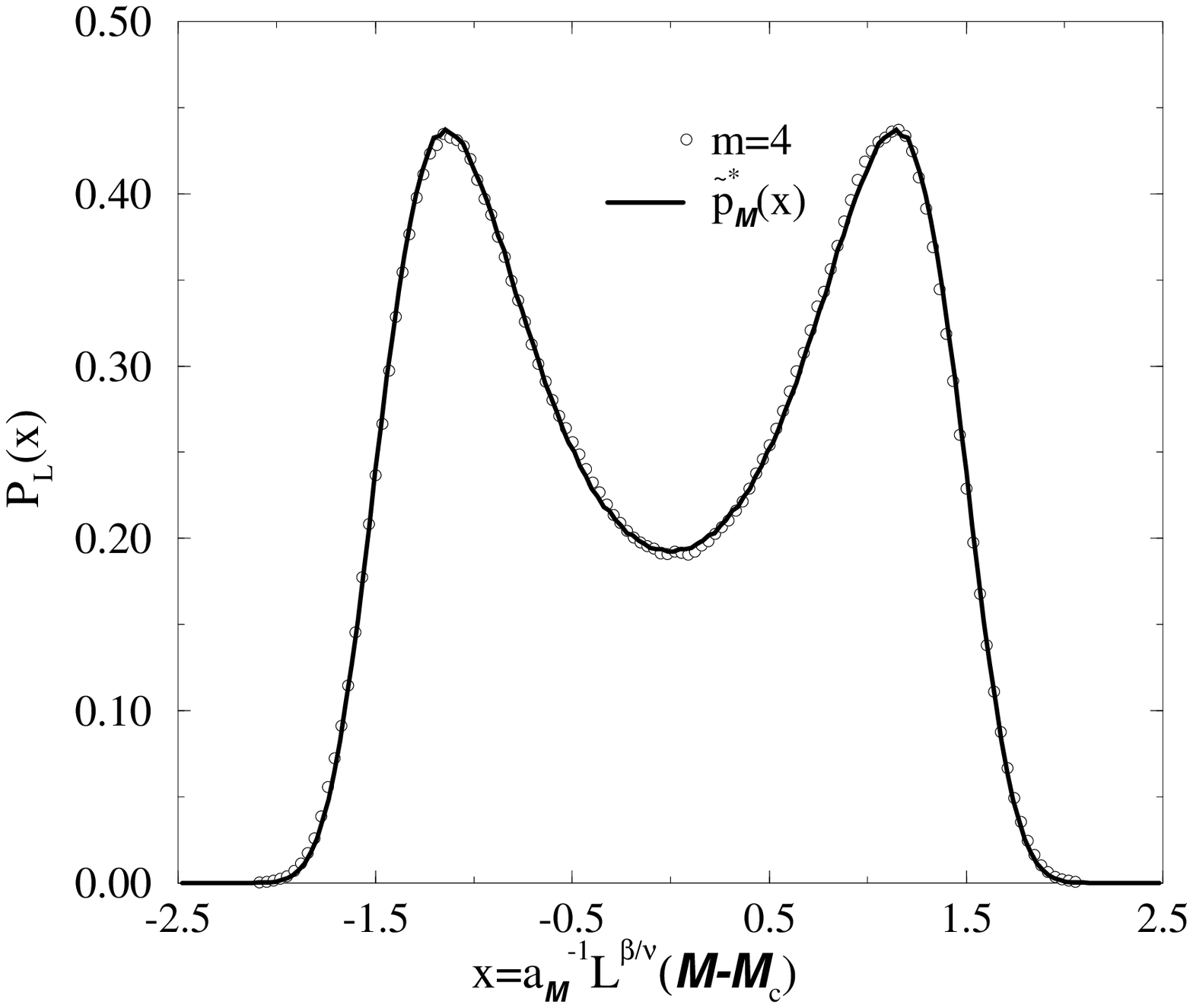}}}

\caption{The measured form of the ordering operator distribution
$p_L({\cal M})$ for the $m=4$ system size at the apparent critical
parameters $T_c^\star=1.1853,\mu_c^\star=-2.7843$. Also shown for
comparison is the universal fixed point ordering operator distribution
$\tilde{p}^\star_{\cal M}(x)$ . The data has been expressed in terms
of the scaling variable $x=a_{\cal M}^{-1}L^{\beta/\nu}({\cal M}-{\cal
M}_c)$, with the value of the non-universal scale factor $a_{\cal
M}^{-1}$ chosen so that the distributions have unit
variance. Statistical errors do not exceed the symbol sizes.}

\label{fig:oM_M4}
\end{figure}

To obtain a preliminary estimate of the critical point parameters, the
universal matching condition for the ordering operator distribution
$p_L({\cal M})$ was invoked. As observed in sections~\ref{sec:intro}
and~\ref{sec:back}, fluid--magnet universality implies that the
critical fluid ordering operator distribution $p_L({\cal M})$, must
match the universal fixed point function $\tilde{p}^\star_{\cal
M}(x)=\int \tilde{p}^\star_{{\cal M},{\cal E}}(x,y) dy$ appropriate to
the Ising universality class. The latter function is identifiable as
the critical magnetisation distribution of the Ising model, the form
of which is independently known from detailed simulation studies of
large Ising lattices \cite{HILFER}. Leaving aside for present the
question of corrections to scaling, the apparent critical point of the
fluid can thus be estimated by tuning the temperature, chemical
potential and field mixing parameter $s$ (within the reweighting
scheme) such that $p_L({\cal M})$ collapses onto
$\tilde{p}^\star_{\cal M}(x)$ . The result of applying this procedure
for the $m=4$ data set is displayed in figure~\ref{fig:oM_M4} where
the data has been expressed in terms of the scaling variable
$x=a_{\cal M}^{-1}L^{\beta/\nu}({\cal M}-{\cal M}_c)$ . The accord
shown corresponds to a choice of the apparent critical parameters
$T_c^\star(L)=1.1853(2), \mu_c^\star(L)=-2.7843(3)$.

Using these estimates of the critical parameters, extensive
simulations were then performed for each of the $5$ system sizes
$m=3-7$ in order to facilitate a full finite-size scaling
analysis. Reweighting was again applied to the resulting histograms to
effect the matching of $p_L({\cal M})$ to $\tilde{p}^\star_{\cal
M}(x)$ thus yielding values of the apparent critical parameters.
Interestingly, however, the apparent critical parameters determined in
this manner were found to be $L$-{\em dependent}. The reason for this
turns out to be significant contributions to the measured histograms
from corrections to scaling, manifest as an $L$-dependent discrepancy
between the critical operator distributions and their limiting Ising
forms. In the case of the ordering operator distribution $p_L({\cal
M})$ , the symmetry of the Ising problem implies that the correction
to scaling function is symmetric in ${\cal M} - \langle {\cal M}
\rangle$. In attempting to implement the matching to
$\tilde{p}^\star_{\cal M}(x)$ we therefore necessarily introduce an
{\em additional} symmetric contribution to $p_L({\cal M})$ associated
with a finite value of the scaling field $\tau$. This latter
contribution has, coincidentally, a functional form that is very
similar to that of the correction to scaling function, a result which
of course make the cancellation of contributions possible. It follows,
therefore, that the magnitude of two contributions must be
approximately equal.

Notwithstanding the added complications that corrections to scaling
engender, it is nevertheless possible to extract accurate estimates of
the infinite-volume critical parameters from the measured histograms.
The key to accomplishing this is the known scaling behaviour of the
corrections to scaling which, (recall equation~\ref{eq:critlimnew}),
die away with increasing system size like $L^{-\theta/\nu}$. Now,
since contributions to $p_L({\cal M})$ from finite values of $\tau$, grow with
system size like $|\tau| L^{1/\nu}$, it follows that implementation of the
matching condition leads to a deviation of the apparent critical
temperature $T_c^\star(L)$ from the true critical temperature
$T_c^\star$ which behaves like

\begin{equation}
T_c^\star(\infty) - T_c^\star(L) \propto L^{-(\theta+1)/\nu}
\end{equation}

\begin{figure}[h]
\vspace*{0.5 in}
\setlength{\epsfxsize}{19cm}
\centerline{\mbox{\epsffile{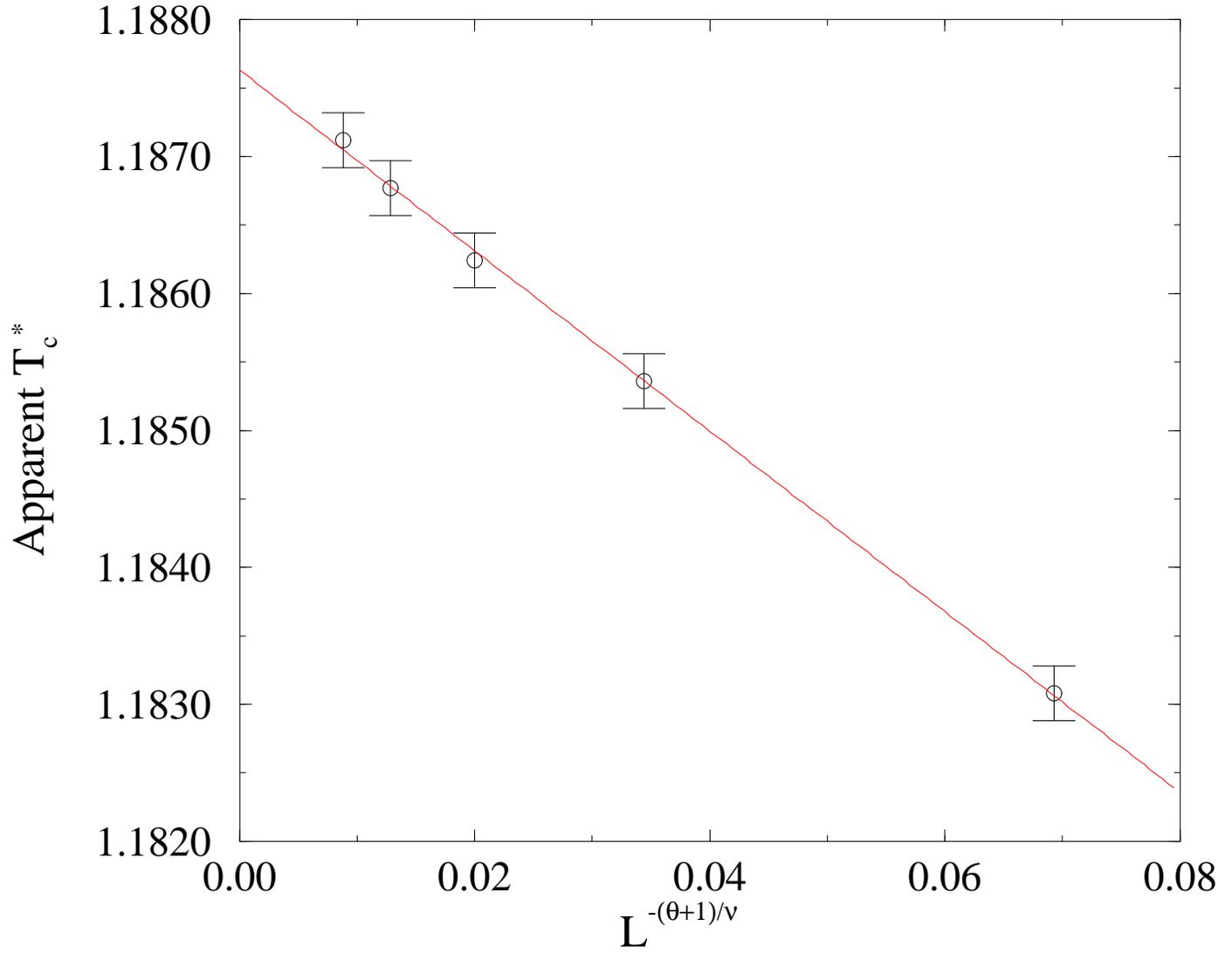}}}

\caption{The apparent reduced critical temperature, (as defined by the
matching condition described in the text), plotted as a function of
$L^{-(\theta+1)/\nu}$, with $\theta=0.54$ and
$\nu=0.629$\protect\cite{FISHER,FERRENBERG1}. The extrapolation of the
least squares fit to infinite volume yields the estimate
$T_c^\star=1.1876(3)$.}

\label{fig:tc_extrap}
\end{figure}

In figure~\ref{fig:tc_extrap} we plot the apparent critical
temperature $T_c^\star (L)$ as a function of
$L^{-(\theta+1)/\nu}$. One observes that the data is indeed well
described by a linear dependence, the least squares extrapolation of
which yields the infinite-volume estimate $T_c^\star=1.1876(3)$. The
associated estimate for the critical chemical potential is
$\mu_c^\star=-2.778(2)$. We note however, that although the
coexistence value of $\mu^\star$ is tightly tied to $T^\star$,
estimates of $\mu_c^\star$ are not directly affected by corrections to
scaling in $p_L({\cal M})$, since the function $\delta p_L({\cal M})
/\delta \mu$ is (to leading order) antisymmetric in ${\cal M}-\langle
{\cal M}\rangle$ \cite{WILDING1}.

\begin{figure}[h]
\vspace*{-0.3 in}
\setlength{\epsfxsize}{11.5cm}
\centerline{\mbox{\epsffile{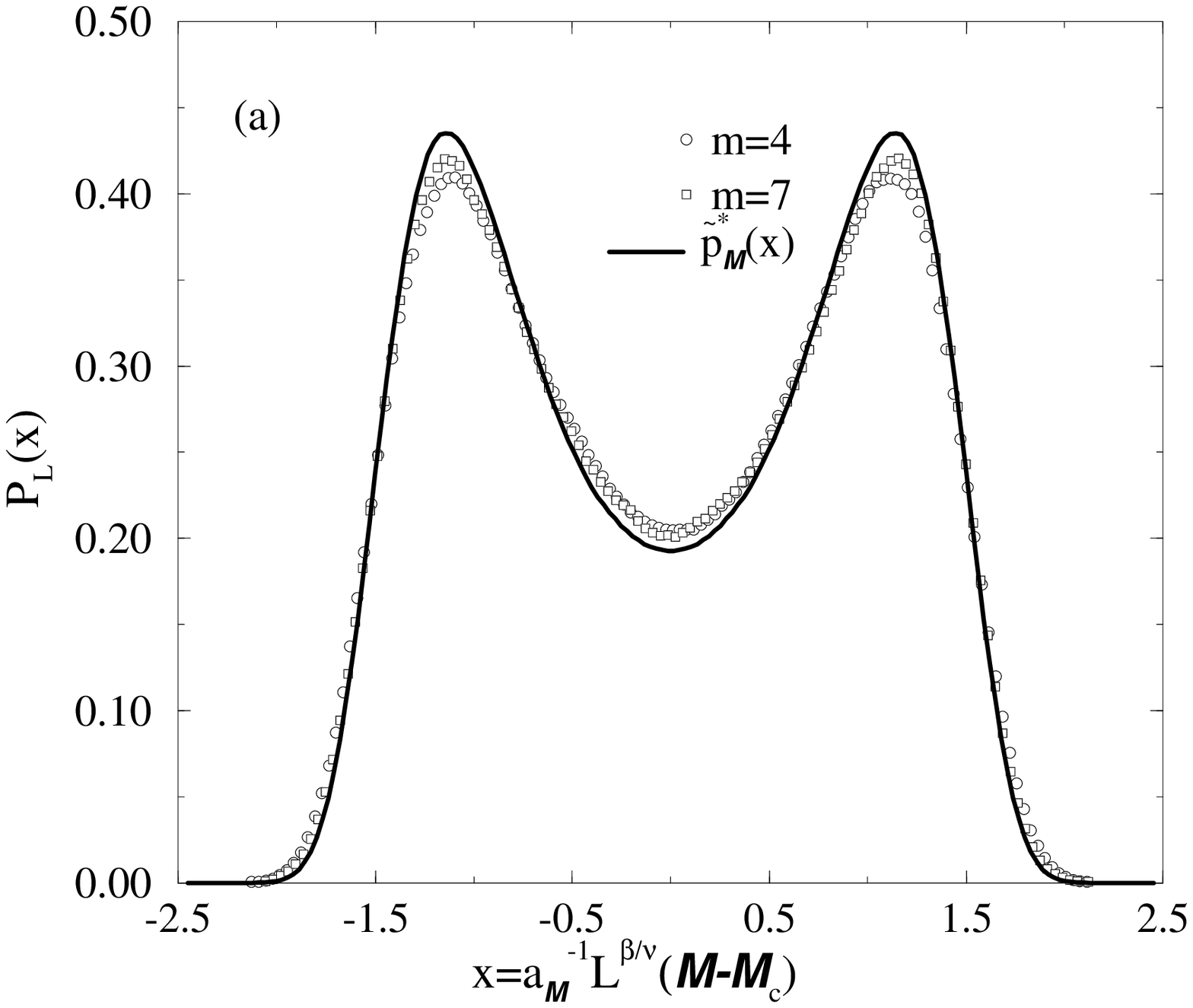}}}
\vspace*{-0.3 in}
\centerline{\mbox{\epsffile{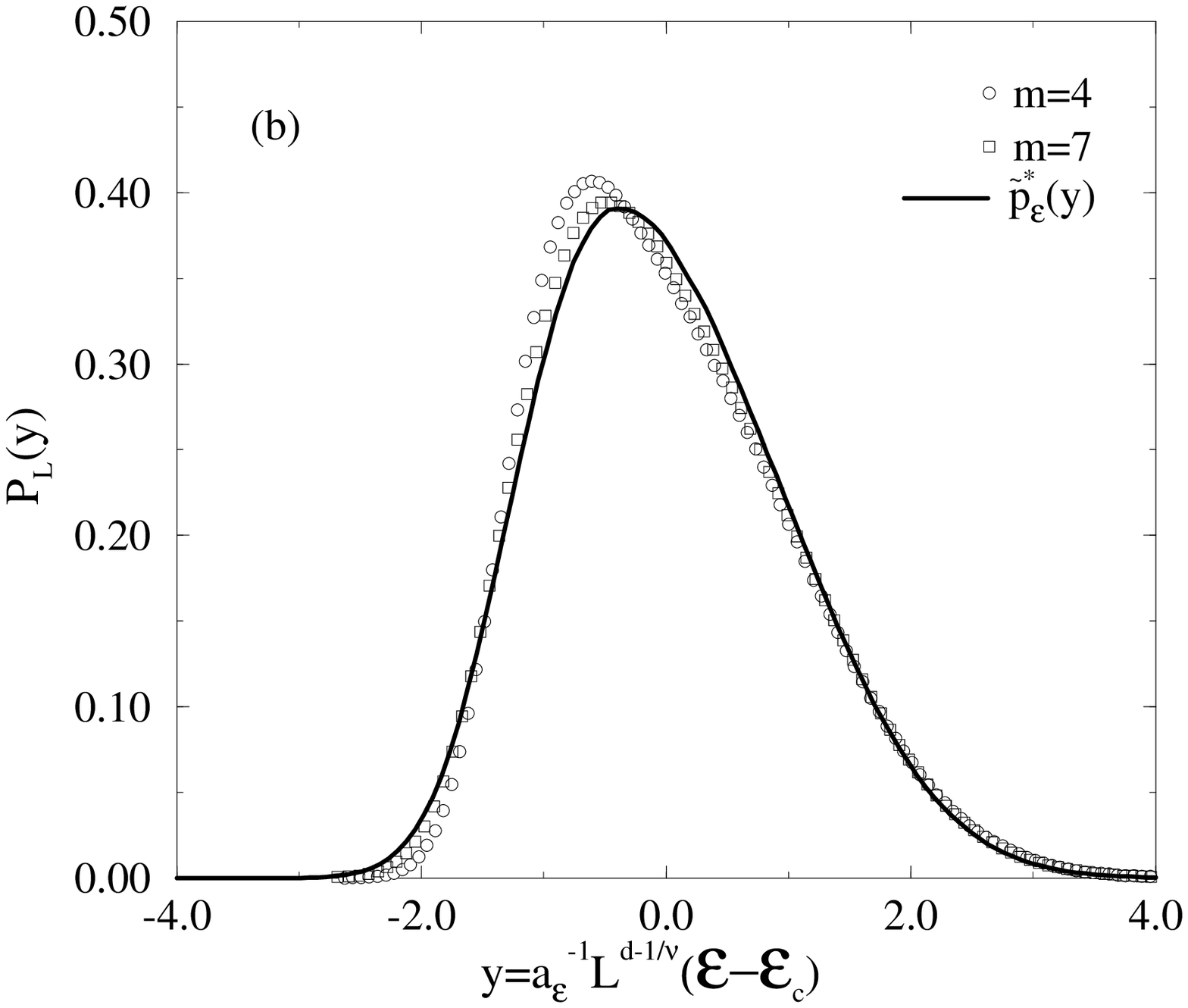}}}

\caption{{\bf (a)} The ordering operator distribution $p_L({\cal M})$
for the two system sizes $m=4$ and $m=7$ at the assigned critical
parameters $T_c^\star , \mu_c^\star$, expressed as function of the
scaling variable $x=a_{\cal M}^{-1}L^{\beta/\nu}({\cal M}-{\cal
M}_c)$.  Also shown (solid line) is the universal fixed point ordering
operator distribution $\tilde{p}^\star_{\cal M}(x)$. {\bf (b)} The
energy operator distribution $p_L({\cal E})$ for the two system sizes
$m=4$ and $m=7$ at $T_c^\star , \mu_c^\star$ ,expressed as a function
of the scaling variable $y=a_{\cal E}^{-1}L^{d-1/\nu}({\cal E}-{\cal
E}_c)$. Also shown (solid line) is the universal fixed point energy
operator distribution $\tilde{p}^\star_{\cal E}(y)$ . In both cases
the values of the non-universal scale factors $a_{\cal E}^{-1}$ or
$a_{\cal M}^{-1}$ have been chosen to yield unit variance.}

\label{fig:oM+oE}
\end{figure}

Having acquired accurate estimates for the infinite volume values of
$T_c^\star$ and $\mu_c^\star$, it is instructive to examine more
closely the size and character of corrections to scaling in the
operator distributions. Addressing first the ordering operator
distribution, we show in figure~\ref{fig:oM+oE}(a), the critical point
form of $p_L({\cal M})$ (expressed in terms of the scaling variable
$x=a_{\cal M}^{-1}L^{\beta/\nu}({\cal M}-{\cal M}_c)$), for the two
system sizes $m=4$ and $m=7$. Also shown is the universal fixed point
function $\tilde{p}^\star_{\cal M}(x)$ appropriate to the 3D Ising
universality class. The corrections to scaling, manifest in the
discrepancy between the fluid finite-size data and the limiting form,
are clearly evident in the figure, especially for the $m=4$ system
size. We note further that their form is qualitatively similar to
those observed in the 2D Ising universality class \cite{NICOLAIDES}.

A similar situation pertains to the energy operator distribution
$p_L({\cal E}) =\int d {\cal M}p_L({\cal M},{\cal E})$.
Figure~\ref{fig:oM+oE}(b) shows the form of this function, together
with the limiting fixed point function $\tilde{p}^\star_{\cal
E}(y)=\int \tilde{p}^\star_{{\cal M},{\cal E}}(x,y) dx$ , identifiable
as the critical energy distribution of the Ising model, and
independently known from detailed Ising model studies
\cite{WILDING3}. The data have all been expressed in terms of the
scaling variable $y=a_{\cal E}^{-1}L^{d-1/\nu}({\cal E}-{\cal
E}_c)$. One observes that in this case, the corrections to scaling are
noticeably larger than for $p_L({\cal M})$ , a fact that presumably
reflects the relative weakness of critical fluctuations in ${\cal E}$
compared to those in ${\cal M}$ .

Turning now to the critical point field mixing parameters, $s$ and
$r$, the values of these quantities were assigned (as described in
detail in reference~\cite{WILDING3}) such as to optimise the mapping
of the critical operator distributions onto their limiting fixed point
forms, (cf. figure~\ref{fig:oM+oE}). The resulting estimates were,
however, found to be slightly $L$-dependent for the smaller system
sizes, an observation that may indicate a finite-size dependence of
the scaling fields themselves \cite{PRIVMAN1}. For the two largest
system sizes this $L$-dependence is, however, small and we estimate
$s=-0.11(1)$, $r=-1.02(1)$.

The measured histograms also serve to furnish estimates of the
exponent ratios $\beta/\nu$ and $1/\nu$ characterising the two
relevant scaling fields $h$ and $\tau$ . These exponent ratios are
accessible via the finite-size scaling behaviour of $p_L({\cal M})$
and $p_L({\cal E})$ at the critical point.  Specifically,
consideration of the scaling form~\ref{eq:critlim} shows that the
typical size of the critical fluctuations in the energy-like operator
will vary with system size like $\delta{\cal E}\sim L^{-(d-1/\nu)}$,
while the typical size of the fluctuations in the ordering operator
vary like $\delta {\cal M}\sim L^{-\beta/\nu}$. Comparison of the
standard deviation of these distributions as a function of system size
thus affords estimates of the appropriate exponent ratios.  In order
to minimise systematic errors resulting from corrections to scaling,
we have performed this comparison only for the two largest system
sizes $m=6$ and $m=7$.  From the measured variance of $p_L({\cal M})$
for these two systems, we find $\beta/\nu=0.521(5)$, an estimate which
compares very favourably with the three dimensional (3D) Ising
estimate \cite{FERRENBERG1} of $\beta/\nu=0.518(7)$. Given though that
no allowances were made for corrections to scaling, the quality of
this accord is perhaps slightly fortuitous.

Carrying out an analogous procedure for $p_L({\cal E})$ yields the
estimate $1/\nu=1.67(7)$, which does not agree to within error with
the 3D Ising estimate $1/\nu=1.5887(4)$. Here, though, we believe that
the bulk of the discrepancy is traceable to the high sensitivity of
$p_L({\cal E})$ with respect to the designation of the field mixing
parameter $r$ implicit in the definition of ${\cal E}$
(cf. equation~\ref{eq:oplinks}). In the presence of sizeable
corrections to scaling, it is somewhat difficult to gauge very
accurately the infinite volume value of $r$ from the mapping of
$p_L({\cal E})$ onto $\tilde{p}^\star_{\cal E}(y)$ . Studies of
significantly larger system sizes than considered here would be
necessary to alleviate this problem.

\begin{figure}[h]
\vspace*{-0.2 in}
\setlength{\epsfxsize}{12.5cm}
\centerline{\mbox{\epsffile{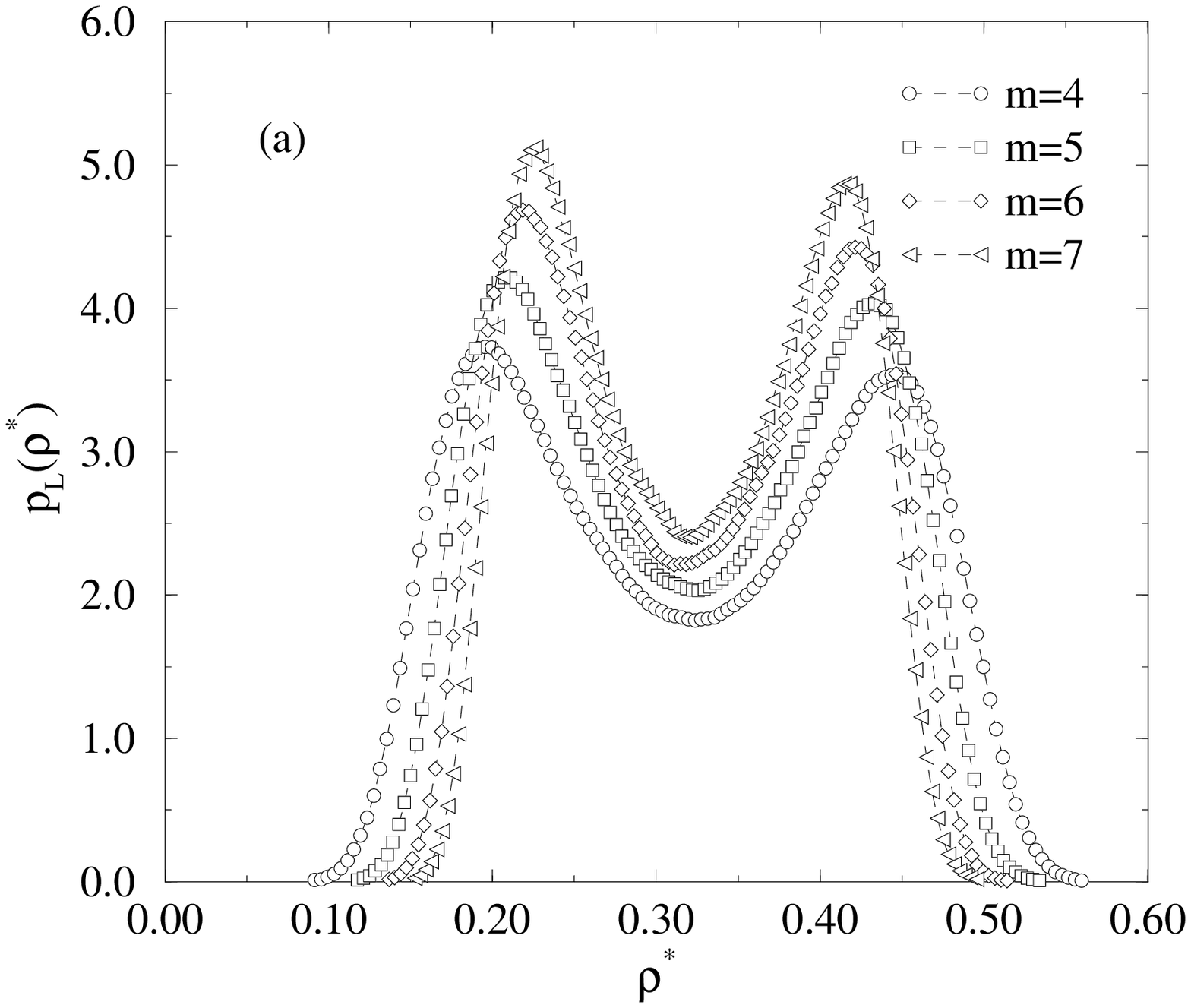}}}
\centerline{\mbox{\epsffile{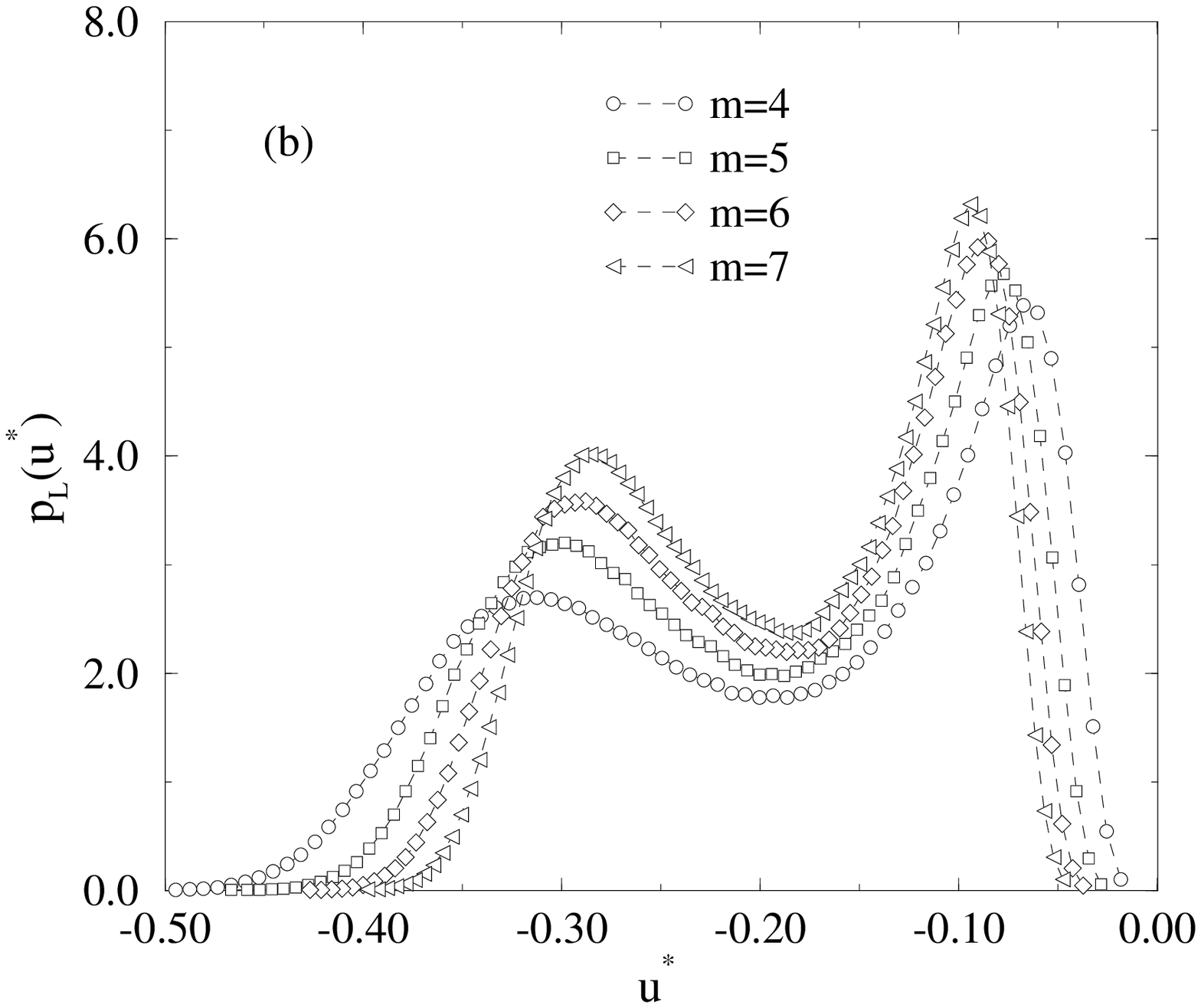}}}

\caption{{\bf (a)} The density distribution at $T_c^\star ,
\mu_c^\star$ for the system sizes $m=4$---$7$, {\bf (b)} The
corresponding energy density distributions. The lines are merely
guides to the eye. Statistical errors do not exceed the symbol sizes.}

\label{fig:rho+en}
\end{figure}

Addressing now the critical density and energy distributions,
figure~\ref{fig:rho+en} shows the measured forms of $p_L(\rho )$ and
$p_L(u)$ at the designated critical parameters.  Clearly these
distributions are to varying degrees asymmetric, a fact which (as
explained in detail in reference \cite{WILDING3}) stems from field
mixing effects. These field mixing contributions, (which are not be
confused with corrections to scaling) die away with increasing $L$ so
that the limiting forms of {\em both} $p_L(\rho )$ and $p_L(u)$ match
the fixed point ordering operator distribution $\tilde{p}^\star_{\cal
M}(x)$.  The approach to this limiting behaviour is indeed quite
evident in figure~\ref{fig:rho+en}.  We note however, that the
limiting form of the fluid critical energy distribution differs from
that of the Ising model where
$\lim_{L\rightarrow\infty}p_L(u)=\tilde{p}^\star_{\cal E}(y)$. This
radical alteration to the limiting behaviour manifests the coupling
that occurs in asymmetric systems between the ordering operator and
energy-like operator fluctuations, the former of which dominate for
large $L$ \protect\cite{WILDING3,MUELLER}. As a consequence one finds
that for critical fluids the specific heat

\begin{equation}
C_v\equiv\ L^d(\langle u^2\rangle-\langle u \rangle^2)/k_BT^2
\sim L^{\gamma/\nu}
\end{equation}
in stark contrast to the behaviour in the Ising model for which
$C_v\sim L^{\alpha/\nu}$. To recapture the Ising behavior it is
instead necessary to consider the fluctations of the fluid energy-like
{\em operator} ${\cal E}$:

\begin{equation}
L^d(\langle {\cal E}^2\rangle-\langle {\cal E} \rangle^2)/k_BT^2
\sim L^{\alpha/\nu}
\end{equation}

\begin{figure}[h]
\vspace*{-0.5 in}
\setlength{\epsfxsize}{12.0cm}
\centerline{\mbox{\epsffile{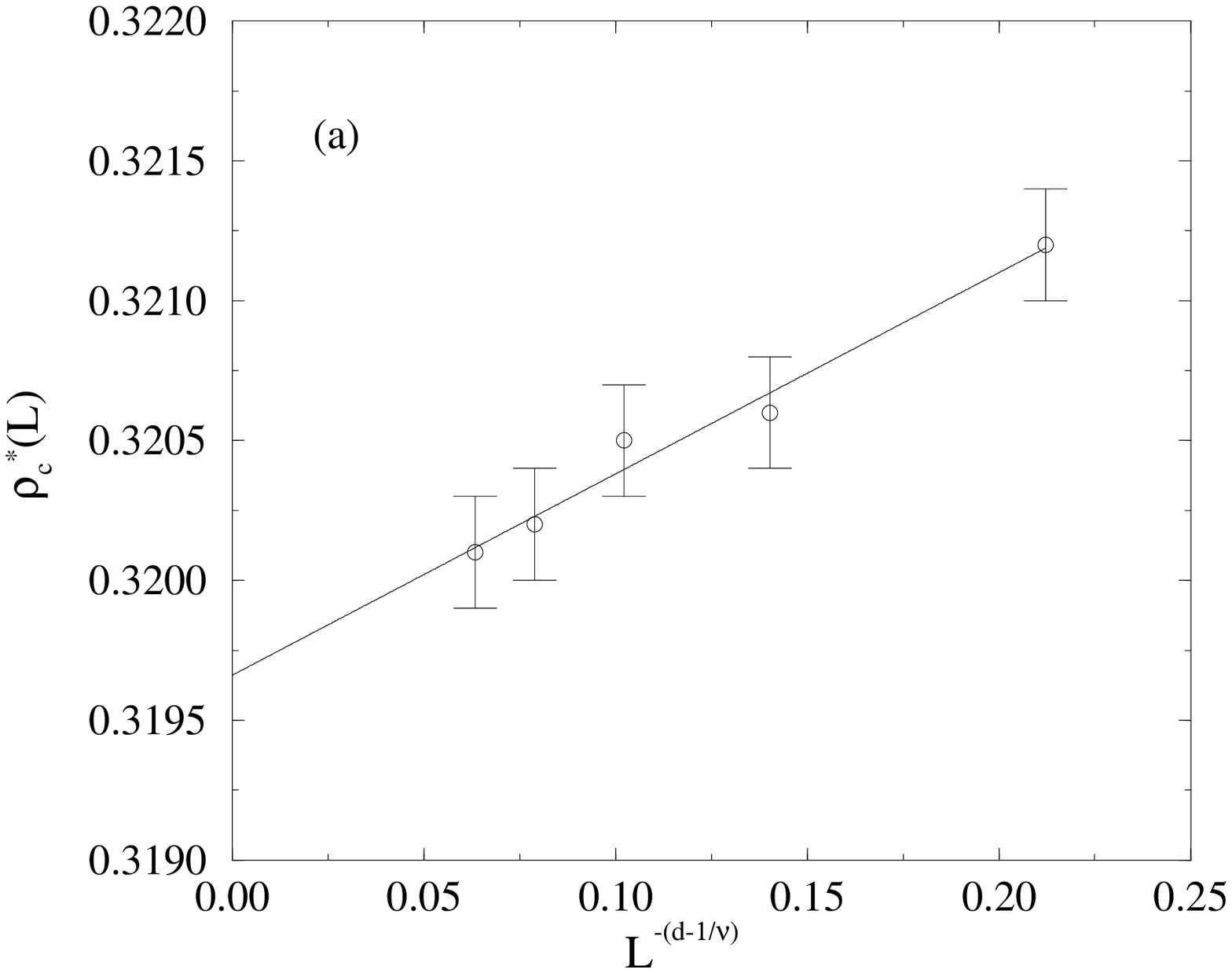}}}
\centerline{\mbox{\epsffile{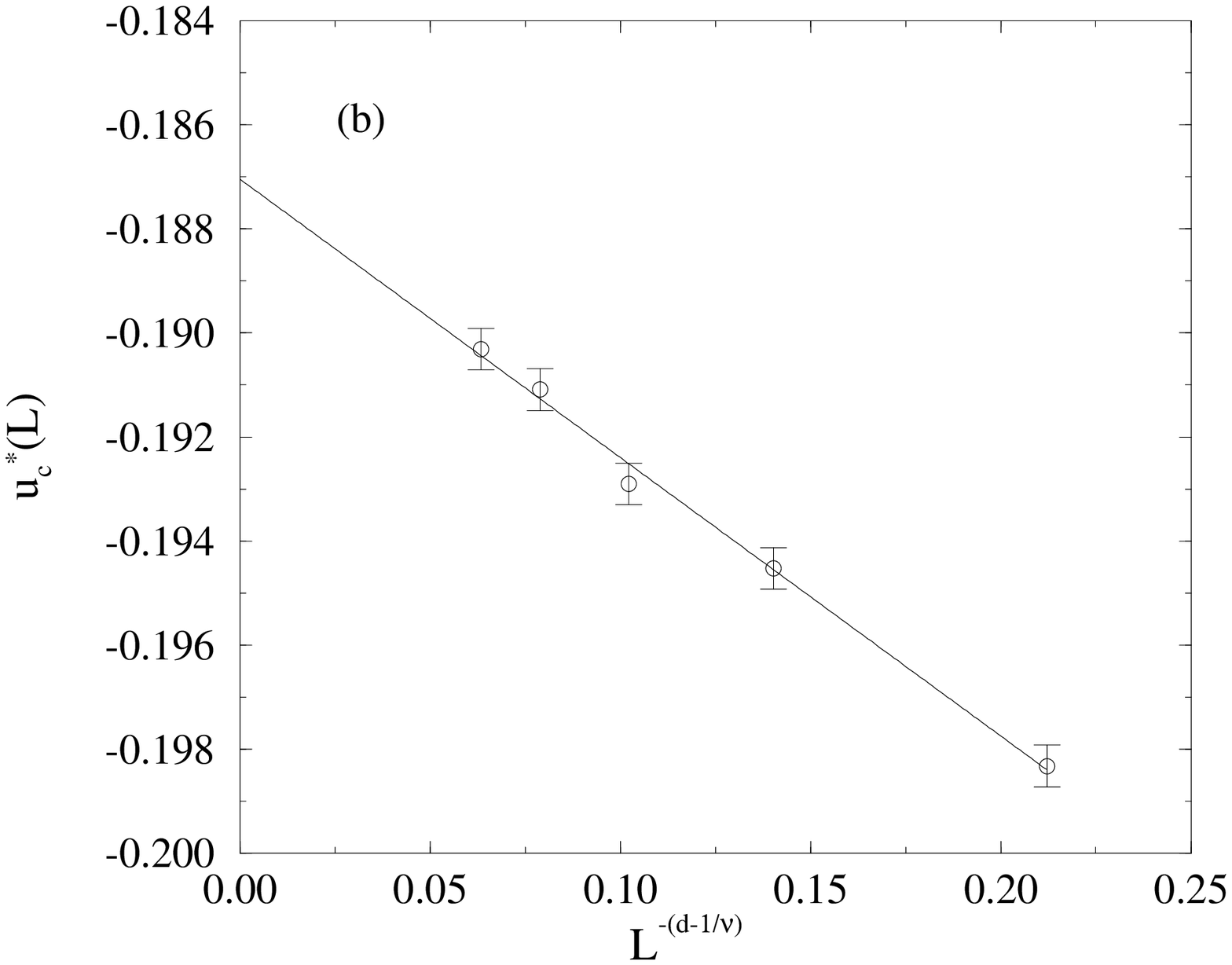}}}

\caption{{\bf (a)} The measured average density
$\langle\rho\rangle_c(L)$ at the designated critical point, expressed
as a function of $L^{-(d-1/\nu)}$. The least-squares fit yields an
infinite volume estimate $\rho_c=0.3197(4)$. {\bf (b)} The measured
average energy density $\langle u\rangle_c(L)$ at the critical point,
expressed as a function of $L^{-(d-1/\nu)}$ . The least-squares fit
yields an infinite volume estimate $u_c=-0.187(2)$. In both cases we
took $1/\nu=1.5887$\protect\cite{FERRENBERG1}. }

\label{fig:rho+en_extrap}
\end{figure}

As a further important consequence of field mixing, it transpires that
measurements of the density and energy distributions at the infinite
volume critical point do not afford direct estimates of the infinite
volume critical density and energy density. This was demonstrated in
reference \cite{WILDING3}, where it was shown that the presence of
field mixing contributions to $p_L(\rho )$ and $p_L(u)$ introduces a
finite-size shift to their average values which behaves like the Ising
energy:

\begin{mathletters}
\begin{eqnarray}
\langle\rho\rangle_c(L)-\langle\rho\rangle_c(\infty ) \sim
L^{-(d-1/\nu)}  \\
\langle u \rangle_c(L)-\langle u \rangle_c(\infty ) \sim L^{-(d-1/\nu)}
\label{eq:limrho+en}
\end{eqnarray}
\end{mathletters}
Thus in order to obtain infinite volume estimates of $\rho_c$ and
$u_c$, it is necessary to perform a finite-size extrapolation of
$\langle \rho \rangle_c$ and $\langle u \rangle_c$ to $L=\infty$. In
figure~\ref{fig:rho+en_extrap} we plot the values of $\langle \rho
\rangle_c$ and $\langle u \rangle_c$, corresponding to the
distributions of figure~\ref{fig:rho+en}, as a function of
$L^{-(d-1/\nu)}$. Although no allowances have been made for
corrections to scaling (the effects of which are certainly much
smaller than those of field mixing), the data exhibit within the
uncertainties, a rather clear linear dependence. Least-squares fits to
the data yield the infinite volume estimates $\rho_c^\star= 0.3197(4),
u_c^\star=-0.187(2)$.

We round off this subsection by summarising our results for the
critical point parameters of the LJ fluid with $r_c=2.5\sigma$:

\begin{eqnarray}
T_c^\star=1.1876(3), & \;\; \mu_c^\star=-2.778(2) \nonumber \\
\rho_c^\star= 0.3197(4), & \;\; u_c^\star=-0.187(2) \nonumber \\
s =-0.11(1), & r=-1.02(1)
\label{eq:critpars}
\end{eqnarray}
A comparison of these estimates with those of previous studies
features in our concluding section.

\subsection{The subcritical coexistence region}

As described in section~\ref{sec:intro}, conventional GCE simulation
studies of the two-phase subcritical region encounter serious problems
due to the large free energy barrier separating the coexisting
phases. This can lead to pronounced metastability effects and
protracted tunneling times between the phases. The multicanonical
ensemble approach \cite{BERG} ameliorates these difficulties by
artificially enhancing the frequency with which a simulation samples
the interfacial configurations of intrinsically low probability. This
enhancement is achieved by sampling not from a simple Boltzmann
distribution with Hamiltonian ${\cal H}(\{{\bf r}\},\rho)$, but from a
modified distribution with effective Hamiltonian ${\cal
H}^\prime(\{{\bf r}\},\rho )={\cal H}(\{{\bf r}\},\rho )+g (\rho )$,
where $g(\rho )$ is a preweighting function the specification of which
is described below.  For the case of the density, the preweighted
distribution takes the form:

\begin{equation}
p^\prime (\rho)=\frac{1}{Z^\prime}\prod_{i=1}^{N=L^d\rho} \left \{ \int
d r_i \right \} e^{-[\Phi(\{{\bf r}\})+\mu L^d\rho +g(\rho)]}
\label{eq:multi}
\end{equation}
where $Z^\prime$ is the multicanonical partition function, which is
defined by equation~\ref{eq:multi}.

If one now chooses the preweighting function such that $g(\rho
)\approx \ln p(\rho)$, where $p(\rho)$ is the desired Boltzmann density
distribution, one readily sees that $p^\prime (\rho) \approx {\rm constant}
\: \forall \rho$. To the extent that this condition is satisfied, the
density thus performs a 1D random walk over its entire domain, thereby
allowing extremely efficient accumulation of the preweighted histogram
$p^\prime(\rho )$. Once this histogram has been obtained, the desired
Boltzmann weighted density distribution is regained as simply

\begin{equation}
p (\rho)=p^\prime (\rho)e^{-g(\rho )}
\end{equation}

Clearly for this approach to succeed, one requires a prior estimate of
the function $p(\rho)$ to use as the preweighting function. But
$p(\rho )$ is, of course, just the function we are trying to find!
While feasible iterative schemes exist for estimating a suitable
weight function \cite{BERG2}, for the purposes of determining the
coexistence curve distributions the task is considerably more
straightforward. Knowledge of a near critical coexistence density
distribution (easily obtainable since the free energy barrier to
tunneling is small near $T_c$) can be used in conjunction with
histogram reweighting and the equal peak-weight criterion \cite{EWR}
to estimate the form of $p_L(\rho )$ for some other chosen point
further down the coexistence line. The extrapolated estimate of the
density distribution may then be used as the weight function in a
multicanonical simulations at this new coexistence state point,
yielding a new coexistence density distribution. The procedure is then
simply repeated: histogram extrapolation of the new distribution being
used to predict the weight function and coexistence parameters
$T^\star ,\mu^\star$ for another state point still deeper into the
subcritical region. In this manner one can systematically track along the
coexistence curve in the space of $\mu^\star$ and $T^\star$, obtaining
at the same time the spectrum of coexistence density distributions.

\begin{figure}[h]
\vspace*{0.5 in}
\setlength{\epsfxsize}{19cm}
\centerline{\mbox{\epsffile{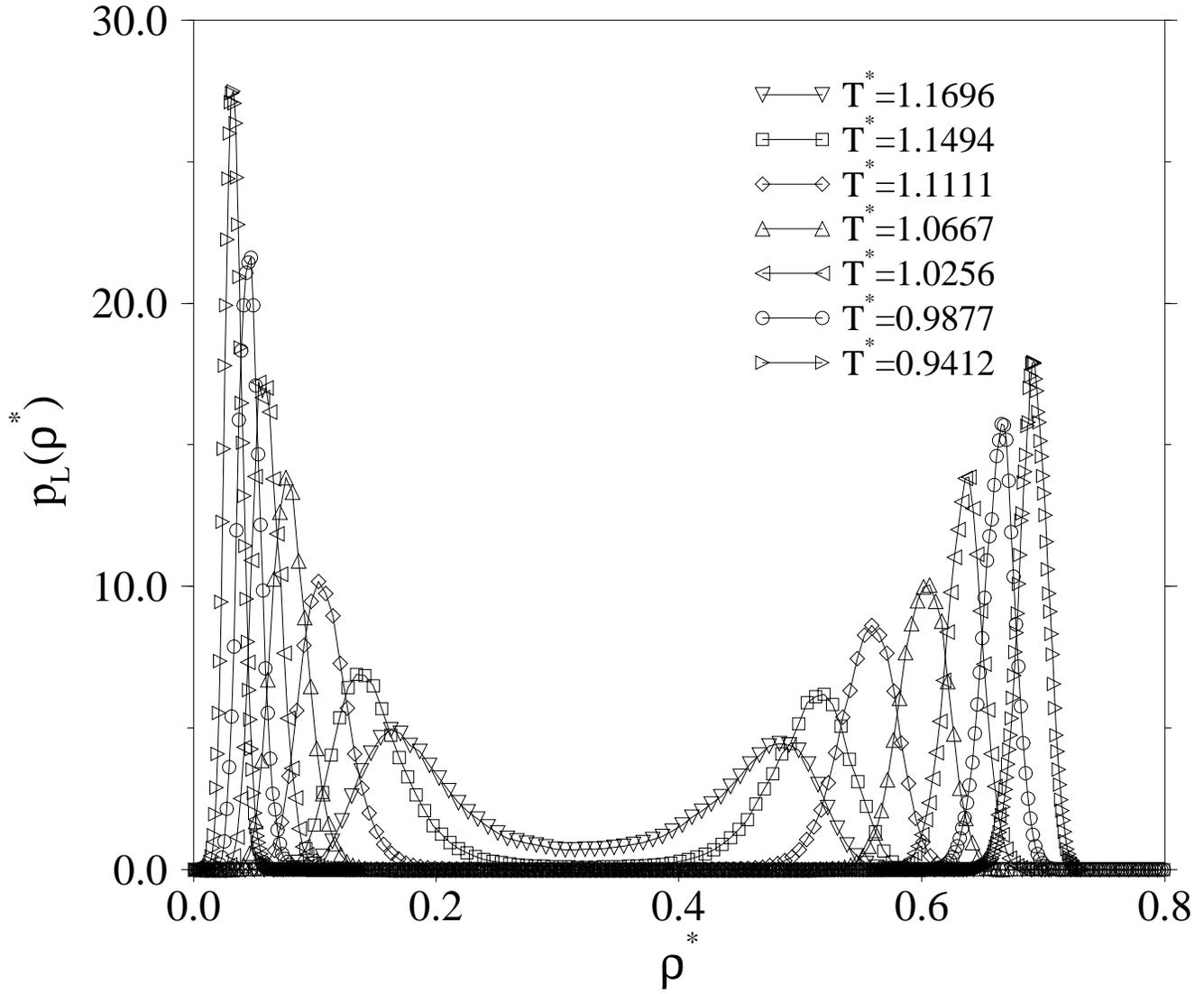}}}

\caption{{\bf (a)} Estimates of the coexistence density distributions
for the $m=4$ systems size, for a range of subcritical temperatures,
obtained as described in the text. The lines are merely guides
to the eye. Statistical errors do not exceed the symbol sizes.}

\label{fig:cxdists}
\end{figure}

We have implemented this strategy for the $m=4$ system, employing the
measured near-critical density distribution as our starting point. It
was found that the histogram reweighting affords reliable
extrapolations over rather a large temperature range: only $7$
multicanonical simulations were required to reach temperature
$T^\star=0.8T^\star_c$. The resulting coexistence density distributions
(corresponding to those temperatures at which the multicanonical
simulations were actually performed) are depicted in
figure~\ref{fig:cxdists}.  We note that for the lowest temperature
studied, $T^\star\approx0.94$, the ratio between the peak and trough heights
of $p_L(\rho )$, is some 30 orders of magnitude!  Such a difference
would, of course, constitute an insurmountable barrier for a
conventional grand canonical simulation.

\begin{figure}[h]
\vspace*{0.5 in}
\setlength{\epsfxsize}{19cm}
\centerline{\mbox{\epsffile{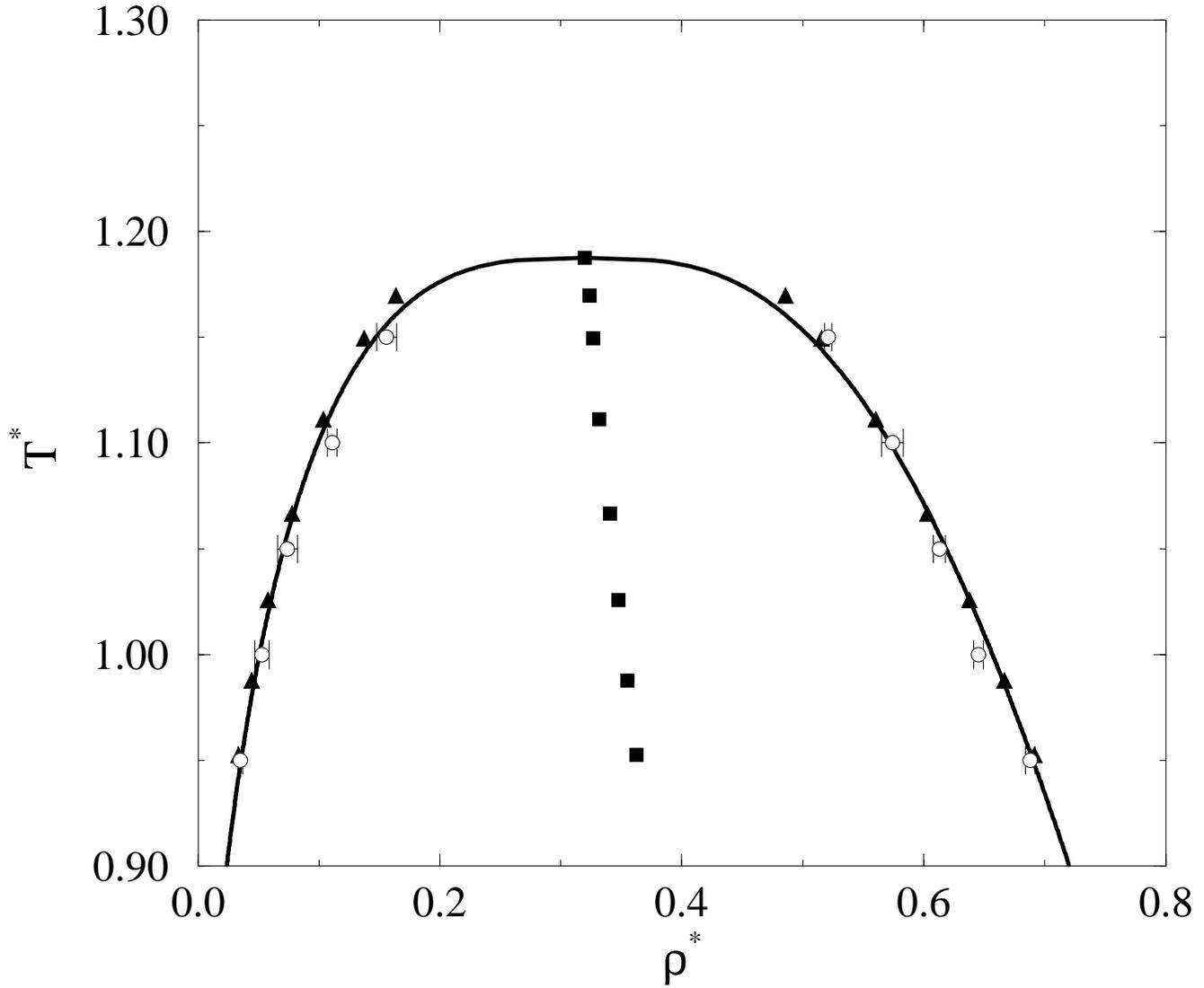}}}

\caption{The peak densities (filled triangles) corresponding to the
distributions of figure~\protect\ref{fig:cxdists}, plotted as a
function of the reduced temperature. The coexistence diameter is also
marked (filled squares). Statistical errors do not exceed the symbol
sizes. Also shown (circles) are the Gibbs ensemble estimates of
Panagiotopolous \protect\cite{PANAGIO3} for a system of size
$L=12\sigma$. The solid line represents a fit through $T_c^\star
,\rho_c^\star$ of the form $\rho_\pm - \rho_c=a| T^\star - T_c^\star
|\pm b|T^\star - T^\star_c|^\beta$, with $a=0.1824(3)$, $b=0.5226(4)$ and
$\beta=0.3258\;$ \protect\cite{FERRENBERG1}.}

\label{fig:cxcurve}
\end{figure}

Away from the immediate vicinity of the critical point (where the
correlation length $\xi \ll L$), the peak positions of the coexistence
density distributions are expected to correspond to the densities of
the infinite volume coexisting phases. This fact can be utilized in
conjunction with the previously determined critical parameters to
estimate the density-temperature phase diagram of the LJ fluid.  In
table~\ref{tab:coexden} we list the peak positions of our measured
density distributions, which are also plotted as a function of
$T^\star$ in figure~\ref{fig:cxcurve}.  For comparison, the GEMC
simulation data of Panagiotopoulos \cite{PANAGIO3} are also
included.

We have attempted to fit our non-critical density data to a
power law of the form

\begin{equation}
\rho_\pm-\rho_c = a|T^\star -T^\star_c|\pm b|T^\star
-T^\star_c|^\beta \label{eq:fit}
\end{equation}
with $T_c^\star$ and $\rho_c^\star$ assigned the values given in
equation~\ref{eq:critpars}, and the order parameter exponent assigned
the Ising estimate $\beta=0.3258 \cite{FERRENBERG1}$ .  The results of
this fit are included in figure~\ref{fig:cxcurve} (solid line) and
correspond to a choice of the critical amplitudes $a=0.1824(3),$
$b=0.5226(4)$. As one observes, the data well away from the critical
point are indeed very well fitted by the assumed form, implying both
that the validity of the scaling form eq.~\ref{eq:fit} extends well
into the subcritical region (a result also observed in many other
simple fluids \cite{PANAGIO2}), and that the coexistence diameter
singularity is undetectably small on the scale of our
measurements. One further sees from figure~\ref{fig:cxcurve} that for
this system size ($m=4$), systematic finite-size effects become
apparent for $T^\star \agt 0.95T_c$. Any power law fit to the density
data that attempted to extrapolate to criticality by including data
points closer to criticality than this, would thus run the risk of
seriously overestimating the critical temperature
\cite{WILDING1,MON,WILDING3}.

\begin{figure}[h]
\vspace*{0.5 in}
\setlength{\epsfxsize}{19cm}
\centerline{\mbox{\epsffile{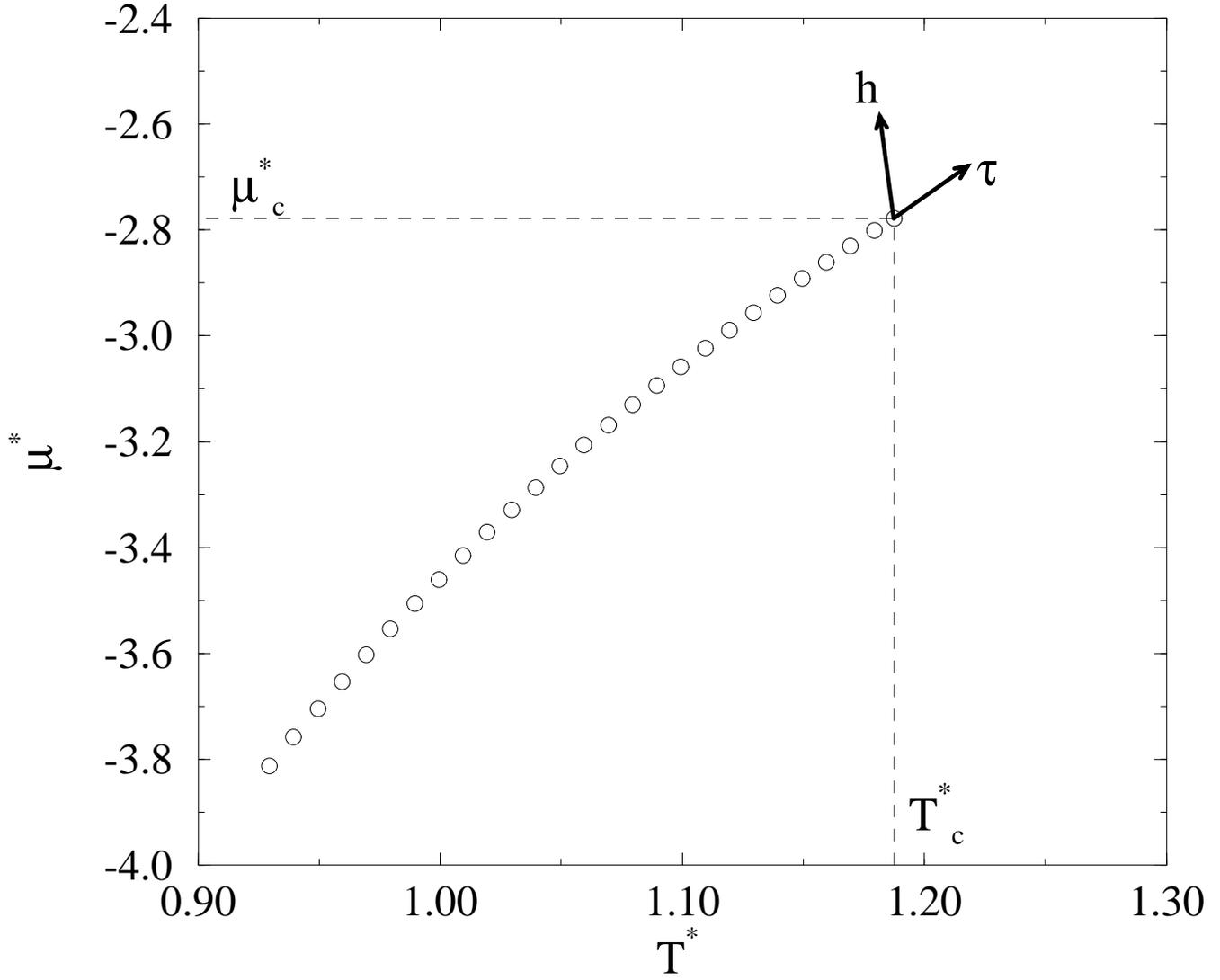}}}

\caption{The line of liquid--vapour phase coexistence in the space of
$\mu^\star$ and $T^\star$, for temperatures in the range $0.95 \leq
T^\star \leq T_c^\star$. The results were obtained by implementing the
equal peak weight criterion for the density distribution \protect\cite{EWR}
in conjunction with the multicanonical simulations and histogram
reweighting. Also shown are the measured directions of the relevant
scaling fields. Statistical errors do not exceed the symbol sizes.}

\label{fig:mu_T}
\end{figure}

Finally, in this section we plot the coexistence curve in the space of
$\mu^\star$ and $T^\star$ as obtained from the multicanonical
simulations. Figure~\ref{fig:mu_T} shows this curve, together with the
estimated critical point and the measured directions of the
relevant scaling fields.

\section{Conclusions}

\label{sec:conc}

In summary, we have employed recently developed mixed-field FSS
techniques and histogram extrapolation methods to obtain highly
precise estimates for the critical point parameters of the truncated
and unshifted LJ fluid with $r_c=2.5\sigma$. Our measurements enable
us to pinpoint the infinite volume critical temperature to within an
uncertainty of $0.03\%$, considerably better than the accuracy of
$1\%$ (or more) typically quoted for other commonly used simulation
techniques.

Two recent studies have also reported values for the critical
parameters of the LJ fluid with $r_c=2.5\sigma$. That of Finn and
Monson \cite{FINN} corrected the equation of state data of Nicolas
{\em et al} \cite{NICOLAS} for the discontinuity at $r_c$ and the
absence of a long tail. Their resulting estimate of the critical
temperature is $T_c^\star=1.23$, which is clearly too high with
respect to our own estimate (equation~\ref{eq:critpars}). Their value
for the critical density $\rho_c^\star=0.32$ does, on the other hand,
agree well with our result, although since no error bars were quoted
it is impossible to tell to what extent the accord is meaningful.

By comparison, the GEMC simulation estimates of Panagiotopoulos
\cite{PANAGIO3} $T_c^\star=1.176(8), \rho_c^\star=0.33(1)$ correspond
rather more closely to our results. In this GEMC study, a power law
fit was made to subcritical coexistence density data, ignoring data points
close to the critical point which are most influenced by finite-size
effects. Although this approach certainly seems to reduce the
systematic overestimate of $T_c^\star$ that can occur in GEMC
simulations when fitting {\em all} the available density data, it is
not clear how much near-critical data should be discarded when the
location of the critical point and the extent of finite-size effects
are not known beforehand. Perhaps as a consequence of this, (as well as
the neglect of corrections to scaling), the error bar on the value of
$T_c^\star$ quoted in \cite{PANAGIO3} does not overlap with ours.

Turning now to the general computational issues raised by the present
study, we have seen the great utility of FSS methods for probing the
critical point region of fluids.  The power of FSS techniques was also
previously demonstrated in a related GCE study of the 2D LJ fluid
\cite{WILDING1}. One significant drawback of this previous
investigation, however, was its high computational cost. Histogram
reweighting was not employed and consequently several long simulations
were required for each $L$, in order to accurately locate the
near-critical coexistence curve and the critical point. This in turn
entailed the use of long runs on high performance parallel
computers. By contrast, use of histogram reweighting in the present
work allowed the study of systems containing up to $5$ times as many
particles as those of the 2D study, while exercising the capabilities
of only a pair of middle-range workstations. We thus believe that the
combined use of FSS methods and histogram reweighting techniques, as
espoused here, brings high precision studies of fluid critical
phenomena within the reach of almost every pocket.

The benefits offered by the use of multicanonical preweighting for
simulations of the subcritical two-phase coexistence region are
similarly impressive. In the previous GCE study of the 2D LJ fluid
\cite{WILDING1}, phase coexistence could not be studied below about
$0.98T_c$, due to the large free energy barrier separating the
coexisting phases. However, by incorporating multicanonical
preweighting into GCE simulations, we have seen that it is possible to
probe much smaller subcritical temperatures with ease \cite{NOTE3}.

Finally we remark that the techniques deployed here are not restricted
to simple fluids, but can be combined with configurational bias Monte
Carlo methods \cite{SIEPMANN,FRENKEL} to facilitate accurate
investigations of phase coexistence and critical phenomena in polymer
systems. We also believe it should be feasible to apply the present
method to systems with long-ranged interactions, e.g. coulombic
fluids, although the computational workload would naturally increase
rapidly with the range of the potential. We hope to report on such
extensions in future work.

\subsection*{Acknowledgements}

The author has benifitted from useful discussions with K. Binder,
A.D. Bruce, D.P. Landau and M.  M\"{u}ller. Helpful correspondence
with M.E. Fisher is also acknowledged. This work was supported by the
Commission of the European Community (ERB CHRX CT-930 351).

\newpage
\narrowtext
\begin{table}
\begin{center}
\caption{The peak densities corresponding to the
coexistence curve distributions depicted in figure~\protect\ref{fig:cxdists}}
\begin{tabular}{|ccc|}
Temperature   &  $\rho_v$  & $\rho_g$   \\
\tableline
1.1696   & 0.1635(10) & 0.4855(10)     \\
1.1494   & 0.1375(9) & 0.5155(9)     \\
1.1111   & 0.1035(9) & 0.5599(9)     \\
1.0667   & 0.0780(9) & 0.6031(9)     \\
1.0256   & 0.0580(9) & 0.6380(9)     \\
0.9877   & 0.0445(9) & 0.6665(9)     \\
0.9412   & 0.0335(8) & 0.6915(8)   \\
%\tableline
\end{tabular}
\label{tab:coexden}
\end{center}

\end{table}

\end{document}